\documentclass[%twocolumn,
aps,nofootinbib,
showpacs,showkeys,preprint
tightenlines,preprintnumbers,] {revtex4}

\usepackage{epsf,epsfig,subfigure,graphicx,amsmath,amssymb %,mathtools
}
\usepackage{color}
\newcommand{\dis}[1]{\begin{equation}\begin{split}#1\end{split}\end{equation}}
\newcommand{\ie}{{\it i.e.~}}

 \newcommand{\dell}{\delta_{\rm PMNS}}
\newcommand{\delq}{\delta_{\rm CKM}}

\newcommand{\Qem}{Q_{\rm em}}

\def\sw0{{$\sin^2\theta_W^0$}}

\def\NDW{N_{\rm DW}}
\newcommand{\Z}{{\bf Z}}

\def\E6{{\rm E_6}}

\def\EE8{{\rm E_8\times E_8'}}

\def\one{{\bf 1}}

\def\five{{\bf 5}}
\def\ten{{\bf 10}}
\def\tenb{{\overline{\bf 10}}}

\def\fiveb{\overline{\bf 5}}

\def\seven{{\bf 7}}

\begin{document}

\draft

\title{  The CKM matrix from anti-SU(7) unification of GUT families}

\author{Jihn E. Kim,$^{(a,b)}$ Doh Young Mo,$^{(c)}$ and Min-Seok Seo$^{(d)}$ }
 
\address{
 $^{(a)\,}$Department of Physics, Kyung Hee University, 26 Gyungheedaero, Dongdaemun-Gu, Seoul 02447, Republic of Korea,  \\
$^{(b)\,}$Center for Axion and Precision Physics Research (IBS),
  291 Daehakro, Yuseong-Gu, Daejeon 305-701, Republic of Korea,\\ 
$^{(c)\,}$Department of Physics, Seoul National University, 1 Gwanakro, Gwanak-Gu, Seoul 151-747, Republic of Korea,\\
$^{(d)\,}$Center for Theoretical Physics of the Universe (IBS),
   Yuseong-Gu, Daejeon  305-811, Republic of Korea 
}

\begin{abstract} 
We estimate the CKM matrix elements in the recently proposed minimal model, anti-SU(7) GUT for the  family unification, $[\,3\,]+2\,[\,2\,]+8\,[\,\bar{1}\,]$+\,(singlets). It is shown that the real angles of the right-handed unitary matrix diagonalizing the mass matrix can be determined to fit the Particle Data Group data. However, the phase in  the right-handed unitary matrix is not constrained very much. We also includes an argument about allocating  the Jarlskog phase in the CKM matrix. Phenomenologically, there are three classes of possible parametrizations, $\delq=\alpha,\beta,$ or $\gamma$ of the unitarity triangle.  For the choice of $\delq=\alpha$, the phase  is close to a maximal one.
 
\keywords{Family unification, Anti-SU(7), CKM matrix, Jarlskog phase,  GUTs}
\end{abstract}
\pacs{12.10.Dm, 11.25.Wx,11.15.Ex}
\maketitle

%%%%%%%%%%%%%%%%%%%%%%%%%%%%%%%%%%%%%%%%%%%%%%%%%%%%
%%%%%%%%%%%%%%%%%%%%%%%%%%%%%%%%%%%%%%%%%%%%%%%%%%%%

\section{Introduction}\label{sec:Introduction}

At present, the unitarity triangle is determined with a very high precision
such that any flavor unification models can be tested against it.  Therefore, we attempt to see whether the recently proposed  unification of grand unification families (UGUTF)  based on anti-SU(7) \cite{KimUGUTF15} is ruled out or not, from the determination \cite{CKMPDG15} of  the Cabibbo-Kobayashi-Maskawa(CKM) matrix elements \cite{Cabibbo63,KM73,Wolfenstein,CKMchau}.  A simple CKM analysis can be performed in the Kim-Seo(KS) parametrization \cite{SeoPRD11} where only complex phase  gives the invariant Jarlskog phase itself \cite{Jarlskog85}. This phase is called the CKM phase $\delq$.

Most family unification models assume a factor group $G_f$ in addition to  the standard model (SM) or grand unification (GUT), where for $ G_f$ continuous symmetries such as SU(2) \cite{Zee79}, SU(3) \cite{KingRoss01}, or U(1)'s \cite{Chun01}, and discrete symmetries such as $S_3$ \cite{S3}, $A_4$ \cite{Ma01}, $\Delta_{96}$ \cite{King14}, $\Z_{12}$ \cite{Dodeca} have been considered. However, a true  unification of families  in the sense that the couplings of the  family symmetry are unified with the three gauge couplings of SM has started with the seminal paper by Georgi \cite{Georgi79}, starting from an SU($N$) GUT \cite{PS73,GG74}. Along this line, a UGUTF based on SU(7)$\times$U(1)$^n$ was suggested  \cite{KimUGUTF15}.
It is derived from string compactification, and contains anti-SU(5) subgroup representations of sixteen chiral fields for one family. These are   ${\ten}_{+1/5}\,(d^c,u,d,N^0),\, \fiveb_{-3/5}\,(d^c,\nu_e,e)$, and $\one_{+5/5}\,(e^+)$ \cite{Barr82,DKN84}.  It is comforting that a plethora of anti-SU(5) or flipped-SU(5) GUTs can be derived in string compactifications \cite{Ellis89,KimKyae07}. 

The anti-SU(7) solution of the  family problem    is to put all fermion representations in 
\begin{equation}
\Psi^{[ABC]}+2\,\Psi^{[AB]}+8\,\Psi_{[A]}+{\rm singlets}\equiv {\bf 35}\oplus  2\times\,{\bf 21}\oplus  8\times\,\overline{\bf 7}+{\one}'{\rm s},\label{eq:UGUTF}
\end{equation} 
where the  indices inside square brackets imply anti-symmetric combinations, and
 %  \dis{
%[1]\equiv \Phi^{[A]}=\begin{pmatrix}
%\alpha_1\\[0.2em] \alpha_2\\[0.2em]  \alpha_3\\[0.2em]  \alpha_4\\[0.2em]  \alpha_5\\[0.2em]  f_6\\ \cdot \\[0.2em]  \cdot\\[0.2em]    \cdot\\[0.3em]  f_N\\[0.2em]
%\end{pmatrix},\hskip 0.5cm
%[2]\equiv \Phi^{[AB]}=\begin{pmatrix}
%0, &\alpha_{12}, &\cdots,&\alpha_{15} & {\Big|}&\epsilon_{16},&\cdots,&\epsilon_{1N}\\[0.2em]
%-\alpha_{12},& 0,&\cdots,&\alpha_{25}&{\Big|}&\epsilon_{26},&\cdots,&\epsilon_{2N}\\ 
%\cdot&  \cdot&  \cdot&  \cdot&{\Big|}&  \cdot& \cdot& \cdot \\ 
%\cdot&  \cdot&  \cdot&   \alpha_{45} &{\Big|}&  \cdot& \cdot& \cdot \\ 
%-\alpha_{15}, &-\alpha_{25},&  %\cdots,&0&{\Big|}&\epsilon_{56},&\cdots,&\epsilon_{5N}\\[0.2em]
%\hline
%-\epsilon_{16},& -\epsilon_{26}\,&\cdots,&-\epsilon_{56}&{\Big|}&0,&\cdots,&\beta_{6N}\\ 
%\cdot&  \cdot&  \cdot&  \cdot&{\Big|}&  \cdot& \cdot& \cdot \\ 
%-\epsilon_{1N},& -\epsilon_{2N}\,&\cdots,&-\epsilon_{5N} &{\Big|}&-\beta_{6N},&\cdots,&0
% \end{pmatrix} \label{eq:FundSUN}
% } 
the bold-faced numbers are the dimensions of representations.
  The color indices are $1,2,3$ and weak indices are $4,5$. With U(1)'s, it is possible to assign the electromagnetic charge  $\Qem=0$ for separating the color and weak charges at the location $[45]$, which is the key point for realizing the doublet-triplet splitting in the GUT BEH multiplets \cite{KimUGUTF15}.  
The merits of  the UGUTF of Ref. (\ref{eq:UGUTF}) are, (i)  it allows the missing partner mechanism naturally based on a suitable $\mu$ parameter \cite{Casas93}, (ii) it is obtained from string compactification, and (iii) it leads to plausible Yukawa couplings. The first   merit  has been already discussed in Ref. \cite{KimUGUTF15}.  The second merit is the following.
The R-parity in SUSY and the Peccei-Quinn symmetry are greatly used for proton longevity and toward a solution of the strong CP problem and cold dark matter \cite{Baer15}. Because of the gravity spoil of such symmetries in general \cite{Gilbert89,Barr92}, discrete gauge symmetries were considered in the bottom approach \cite{KraussWilczek,Ibanez92}. It can be a discrete subgroup of some gauge group.  In the top-down approach, such as in models from string compactification,  the resulting approximate discrete and global symmetries are automatically allowed since string theory describes gravity without such problems \cite{KimNilles14, KimPRL13}.  
 
In this paper, we focus on the third merit by adopting the spectra obtained in Ref. \cite{KimUGUTF15}, and explicitly calculate the CKM matrix. Here, we do not use the full description of Yukawa couplings dictated from string theory \cite{Hamidi87}, but use just the supergravity couplings including non-renormalizable terms\footnote{String calculation of all non-renormaliable couplings are not available at present. See, for an attempt, Ref. \cite{ChoiKobayashi08}.} suppressed by the string scale, $M_s$. Thus, every nonrenormalizable term introduces an undetermined coefficient of O(1). Here, we reduce the number of couplings,  using the $\Z_{12-I}$ discrete symmetry implied from its origin of $\Z_{12-I}$ orbifold compactification \cite{KimUGUTF15}.
  
  %%%%%%%%%%%%%%%%%%%
  \section{Some comments related to the Jarlskog determinant}
  
It is known that   $\delq\approx 90^{\rm o}$ in the KS parametrization \cite{KimJKPS15}. The Particle Data Group(PDG) compilation gives $\delq=(85.4^{+3.9}_{-3.8})^{\rm o}$, \ie our $\delq$ is their $\alpha_{\rm PDG}$ \cite{CKMPDG15}. We consider this as a maximal phase with the prescribed real angles. 
The Jarlskog determinant $J$ is the area of the Jarlskog parallelogram which has two angles whose sum is $\pi$. The area of the parallelogram has the form: $(\it combination\,of\,real\,angles)\cdot\sin\delq$. So, the Jarlskog phase can be taken as $\delq$ or $\pi-\delq$.  Let us define `Jarlskog triangle' by cutting the parallelogram along a diagonal line, and the Jarlskog invariant phase is the angle opposite to the cutted line. One crucial question is whether the Jarlskog phase is parametrization-independent or not. This is because the Jarlskog determinant, \ie the area, can be made the same in different parametrization schemes by appropriately changing  $(\it combination\,of\,real\,angles)$ and $\sin\delq$. We argue that there are only three classes of the CKM parametrizations from length sides of O($\lambda^3$) unitarity triangle. From the unitarity triangle of $B_s$ meson decay with O$(\lambda^3)$ lengths, there are three angles $\alpha,\beta$ and $\gamma$, and we can define three classes of parallelograms with the same area. Since there are six different unitarity triangles, three {\it vertical cases} in choosing two columns and  three {\it horizontal cases} in choosing two rows, the total number of possibilities is 18.  
Out of these 18 angles, 4 angles (having $\delq\simeq 0$) with side lengths of O$(\lambda)$ and  O$(\lambda^2)$ from horizontal and vertical cases are phenomenologically excluded. Furthermore, the invariant phase appears in all six triangles.  Since the unitarity triangle of $B_s$ meson decay is known rather accurately, only three angles are suitable for  $\delq$.  The KS parametrization uses $\alpha$ of  the unitarity triangle of $B_s$ meson decay as $\delq$ while the Chau-Keung-Maiani (CK) parametrization \cite{CKMchau} uses $\gamma$ as $\delq$. Since $\alpha\simeq 90^{\rm o}$, it is minimal (also, see below)   to adopt the KS parametrization since there can be one Jarlskog phase $\frac{\pi}{2}$. In the other classes, there are two Jarlskog phases, $\gamma$ and $\pi-\gamma$, or $\beta$ and $\pi-\beta$.  
 
  If the phase is parametrization-dependent, it is not so important to try to determine very accurately $\alpha, \beta, \gamma$ in the unitarity triangle of $B_s$ meson decay in Particle Data Book \cite{CKMPDG15}. Here, we argue that the CKM phase $\delq$ is scheme independent up to three classes. Assume that the weak CP violation is introduced spontaneously \cite{TDLee73} by a complex vacuum expectation value of the standard model (SM) singlet $X$ \cite{KimmaxCP}. Suppose
the phase  of $\langle X\rangle$ is $2\pi n/\NDW$ where $n$ and $\NDW$ do not have a common divisor. Thus, the vacuum has $\NDW$ different domains which are separated by domain walls  \cite{Kibble}. Depending on the value of $\delq$, we can similarly define the domain wall number of the CKM matrix, $N_{\rm CKM}$.  Let the phase $\delta$ of the SM singlet $X$ vary continuously from $0$ to $2\pi$. Along this variation, one passes through the different domains of number $\NDW$.  Now, suppose we perform weak CP violation experiments looking at the Jarlskog phase. Observe that $\delq$ must be proportional to the phase of  $\langle X\rangle$ since there will be no CP violation if  $\langle X\rangle$ is real.   In the same domain, measurements on the weak CP phase must be identical. So, we obtain $N_{\rm CKM}\le \NDW$. In addition, in the two adjacent domains, measurements on weak CP phase must be different, leading to  $N_{\rm CKM}\ge \NDW$. Thus, we obtain $N_{\rm CKM}= \NDW$. In this Gedanken experiment with spontaneous CP violation, $|\delq|$ is the magnitude of the phase of the VEV $\langle X\rangle$. So, it must be scheme independent, up to three classes, since in any CKM parametrization  the VEV of fundamental field $X$ is not introduced. In this argument, it is better to use the parametrization scheme where the phase of  $\langle X\rangle$ sits at origin, regardless of the product of the combinations of real angles.  Namely, the phase $\delq$ has an invariant meaning, up to three classes. It is a topological argument and may be applicable for the case of complex Yukawa couplings also because one can mimick the complex Yukawa couplings by VEVs of SM singlets. Thus, the first result from the Jarlskog determinant is that the phase $\delq$ has a  parametrization-independent meaning, up to three classes. The second result is that the product of the lengths of two sides enclosing the Jarlskog phase $\delq$ is invariant. 
  
The third aspect is the following. Firstly, as an illustration, consider the Kobayashi-Maskawa(KM) parametrization \cite{KM73}. Its determinant is not 1, but $ -e^{i \delta' }$. A proper redefinition, making the determinant real and rotating all six Jarlskog triangles without changing the shapes, is to multiply  $ e^{i(\pi-\delta')/3}$ to every element. It is equivalent to multiplying $e^{i\delta_0}$ ($\delta_0=(\pi-\delta')/3$) to $\bar{u}_L,\bar{c}_L,$ and $\bar{t}_L$ fields such that the newly defined primed fields are $\bar{u}_L=e^{i\delta_0}\bar{u}_L',\bar{c}_L=e^{i\delta_0}\bar{c}_L',$ and $\bar{t}_L=e^{i\delta_0}\bar{t}_L'$. Then, obviously the shapes of all six triangles are not changed. But this introduces a factor $e^{i\pi/3}$ in every elements. To keep the shapes of at least three {\it vertical} Jarlskog triangles, including the familiar one in the PDG book,  multiply  diag.\,$(1,1,-e^{-i\delta'})$ on the right-hand side of the KM matrix, leading to
 \dis{ 
V'_{\rm KM}=\begin{pmatrix}
c_1 ,& -s_1c_3 ,& e^{-i\delta'}s_1s_3  \\
c_2s_1 ,&-e^{i\delta'}s_2s_3+c_1c_2c_3, & -s_2c_3-e^{-i\delta'}c_1c_2s_3  \\
s_1s_2, & c_2s_3e^{i\delta'}+c_1s_2c_3  ,& c_2c_3-e^{-i\delta'}c_1s_2s_3  
\end{pmatrix}
}
from which we obtain
\dis{&\alpha = {\rm Arg.}\Big(-\frac{V_{td}V_{tb}^*}{V_{ud}V_{ub}^*}\Big)={\rm Arg.}\Big(\sin\theta_2(-e^{-i\delta'}\cos\theta_2\cot\theta_3\sec\theta_1
+\sin\theta_2)\Big),
\\
&\beta = {\rm Arg.}\Big(-\frac{V_{cd}V_{cb}^*}{V_{td}V_{tb}^*}\Big)={\rm Arg.}\Big(\frac{\cos\theta_2(\cos\theta_3
+e^{i\delta'}\cos\theta_1\cot\theta_2
\sin\theta_3)}{\cos\theta_2\cos\theta_3-e^{i\delta'}\cos\theta_1\sin\theta_2\sin
\theta_3}\Big),
\\
&\gamma = {\rm Arg.}\Big(-\frac{V_{ud}V_{ub}^*}{V_{cd}V_{cb}^*}\Big)={\rm Arg.}\Big(\frac{\cos\theta_1\sec\theta_2
\sin\theta_3}{e^{-i\delta'}\cos\theta_3\sin\theta_2
+\cos\theta_1\cos\theta_2\sin\theta_3}\Big).
\label{eq:alphagamma}
}
Note that $\alpha\simeq \pi-\delta'$, \ie the KM parametrization uses $\alpha$ of PDG as $\delq$.\footnote{Note that $\delq$ is defined $\alpha$ or $\pi-\alpha$, depending on the cutted diagonal line.}  

It is very useful if the CKM matrix itself contains the invariant phase $\delq$ in a visible manner. $J$ is always arising with O($\lambda^3$) multiplied. Since the (13) and (31) elements of $V_{\rm CKM}$ are already O($\lambda^3$) \cite{Wolfenstein}, it is convenient for the phase $e^{i\delq}$ to appear either in the (31) element or in the (13) element with one row or one column real. Since the (22) element is an almost real O(1) constant, it will lead to $J={\rm Im\,}V_{31}^*V_{22}^*V_{13}^* =O(\lambda^6)\sin\delq$ \cite{KimJKPS15}.\footnote{This form is true in any parametrization with Det.\,$V_{\rm CKM}=1$.
}  
It is convenient to make the first row real, \ie the (13) element real. Then, the denominator $V_{ud}V^*_{ub}$ in Eq. (\ref{eq:alphagamma}) is real and the Jarlskog triangle has one side on $x$-axis. The angle at the origin is $\delq$. On the other hand, the CK parametrization \cite{CKMchau} has real values for both the first row and first column, and its determinant $=1$. Thus, $J $ must be contributed from the phase in the (22) element. The $V_{(cd)}\cdot V^*_{(cb)}$ component (for the (22) element, or the $(cc)$ element in the mass eigenstate bases) appears for $\beta$ and  for $\gamma$  in Eq. (\ref{eq:alphagamma}).
For the one in the numerator, \ie in $\beta$, $V_{tb}^*$ in the denominator is also complex, and  $V_{(cd)}\cdot V^*_{(cb)}$ alone cannot determine $\delq$. On the other hand, the one in the numerator, \ie in $\gamma$, the numerator is real, and  $V_{(cd)}\cdot V^*_{(cb)}$ alone  determines $\delq$. Thus, $\delq$ is $\gamma$ in the CK parametrization. We can generalize this statement. Let us use the parametrizations such that the large components of the diagonal elements are real. Then, if the first row or first column is real, $\delq=\alpha$. If both the first row and first column are real, then $\delq=\gamma$. To have $\beta$ as $\delq$, we need that the (22) element contains a large impaginary part.   In this analysis, it was useful to remember the formula of Ref. \cite{KimJKPS15}: $J ={\rm Im\,} V^*_{31}V^*_{22}V^*_{13}$.

The invariant Jarlskog phase appears in all Jarlskog triangles, not necessarily at the origin.
 Let us take, as an illustration purpose, $\alpha\simeq90^{\rm o}=\frac{2\pi}{4}, \beta\simeq 22.5^{\rm o}=\frac{2\pi}{16}$, and $\gamma\simeq 67.5^{\rm o}=\frac{2\pi}{16} \times 3$ which are within the experimental bounds. If these phases appear from some $\Z_N$ symmetry, we can choose three kinds of $N$ depending on which angle is used for $\delq$. If $\alpha, \beta$, and $\gamma$ are used for $\delq$, $\NDW$ of $\langle X\rangle$ must be 4, 16, and 16, respectively.    In this paper we use the  KS parametrization  \cite{SeoPRD11,KimJKPS15}, which is a kind of minimal one,
 \dis{
 V_{\rm KS}&=\left(
\begin{array}{ccc}
c_1 & s_1c_3 & s_1s_3  \\
-c_2s_1 & e^{-i\delq}s_2s_3+c_1c_2c_3 & -e^{-i\delq}s_2c_3+c_1c_2s_3  \\
-e^{i\delq}s_1s_2 & -c_2s_3+c_1s_2c_3 e^{i\delq} & c_2c_3+c_1s_2s_3e^{i\delq} \\
\end{array}\right)\label{eq:KSmatrix}
. } 
Note that $J$ is  given as, $J_{\rm KS}={\rm Im\,} V^*_{31}V^*_{22}V^*_{13}=c_1c_2c_3
 s_1^2s_2s_3\sin\alpha=O(\lambda^6-
 \lambda^7)$ in the KS parametrization and $J_{\rm CK}={\rm Im\,} V^*_{31}V^*_{22}V^*_{13}= c_{12} c_{13}^2c_{23}s_{12} s_{13} s_{23}   \sin\gamma=O(\lambda^6-
 \lambda^7)$   in the CK parametrization. If the Cabibbo angle $\theta_C=s_1c_3= s_{12}c_{13}$ is fixed, $J/\sin\theta_C= 
c_1c_2s_1s_2s_3\sin\alpha=c_{12} c_{13} c_{23}  s_{13} s_{23}   \sin\gamma $. For a numerical study, we can choose a {\it vertical} Jarlskog triangle of the first and second columns, where two O($\lambda$) side lengths are $|c_1c_3s_1|, |c_2s_1(c_1c_2c_3+s_2s_3 e^{-i\alpha})|$, and an O($\lambda^5$) side length is $ e^{i\alpha}s_1s_2|(c_1c_3s_2-c_2s_3e^{-i\alpha})| $ with the phase explicitly written  for the  O($\lambda^5$) side to rotate it freely. The corrected area depending on $\theta_2,\theta_3$ and $\alpha$ is $J/c_1\sin^2\theta_C=\frac12\sin(2\theta_2) \tan(\theta_3)\sin\alpha$. For given $\sin 2\theta_2$ and $\tan\theta_3$,  we can rotate $\alpha$ to $90^{\rm o}$ to obtain the largest $\delq$ since in our choice of  $\alpha\sim 90^{\rm o}$ is allowed. We cannot give this argument for $\delq=\gamma$, where $\gamma$ is far from $90^{\rm o}$.

It is pointed out that if $\delq=\pm\dell$ is empirically proved then the idea of spontaneous CP violation {\`a la} Froggatt and Nielsen with a UGUTF makes sense \cite{KimNam15}. In this case, the value $\dell$ will choose one class of the CKM parametrizations we discussed here.

%%%%%%%%%%%%%%%%%%%%%%%%%%%%%%%%%%%%%
\section{Yukawa couplings and masses}\label{subsec:Yukawa}

%%%%%%%%%%%%%%%%%%%%%%%%%%%%%%%%%%%%%
\subsection{U(1) charges in anti-SU(7)}\label{subsec:Uones}

To check the Yukawa couplings, it is useful to have U(1) charges in the anti-SU(7) model. For completenes, therefore, we list them. For the fundamental representation \seven, the U(1) charges belonging to SU(5) and SU(7) are defined as
\dis{
&X_5=\left( \frac{2}{30},\, \frac{2}{30},\, \frac{2}{30},\, \frac{-3}{30},\, \frac{-3}{30},\,0,\,0\right) \\
&Z_7=\left( \frac{-2}{7},\, \frac{-2}{7},\, \frac{-2}{7},\, \frac{-2}{7},\, \frac{-2}{7},\, \frac{ 5}{7},\, \frac{ 5}{7} \right) \label{eq:Uones}
}
The extra U(1) charge beyond SU(7) is
\dis{
Z=\left( \frac{-5}{7},\, \frac{-5}{7},\, \frac{-5}{7},\, \frac{-5}{7},\, \frac{-5}{7},\, \frac{-5}{7},\, \frac{-5}{7} \right) .\label{eq:SU7U1Z}
}
For the matter \seven, therefore, we represent it as
${\bf 7}_{-5/7}$.
The electroweak hypercharge $Y$ of the SM and the U(1) charge $X$ of the flipped-SU(5) are defined as
\dis{
&Y=\left( \frac{-1}{3},\, \frac{-1}{3},\, \frac{-1}{3},\, \frac{ 1}{2},\,  \frac{1}{2},\,0,\,0\right)=X_5+X, \\
&X=\left( \frac{18}{30},\, \frac{18}{30},\, \frac{18}{30},\, \frac{18}{30},\, \frac{18}{30},\,0,\,0\right) 
=-\frac{3}{5}(Z_7+Z),
\label{eq:YandX}
}
When $\bf 21$ branches to SU(5) representations  $\ten,\,2\cdot{\five}$, and $\one$, the SM U(1) charges are required to be the familiar ones, determining subscripts $a,b,c$ in the following , 
\dis{
& \left(\ten_a;\,\five_b,\,\five_b;\,\one_c  \right)= \left(\frac{1}{3}(d^c),\,\frac{1}{6}(q),\,0(N);\,\five_a,\,\five_a,\,\one_b  \right)\to a=\frac{1}{5}\,(\sim \frac{6}{5}),  ~b=\frac{3}{5},~c=0, \label{eq:XAB}
}
where we used Eqs. (\ref{eq:Uones}) and (\ref{eq:YandX}) and used quantum numbers of  ${\bf 21}=\Psi^{[AB]} $. When ${\bf 35}$ branches to SU(5) representations as $\tenb,\,2\cdot\ten,$ and $ {\five}$, similarly  subscripts $d,e,f$ in the following are determined as
\dis{
 \left(\tenb_d;\,\ten_e,\,\ten_e;\,\five_f  \right)&= \left(\frac{-1}{3} ,\,\frac{-1}{6} ,\,0(\overline{N})\, ; \,2\left[\frac{1}{3}(d^c),\,\frac{1}{6}(q),\,0(N)\right];\,\five_g  \right)\\
&\to d=\frac{-1}{5}\,(\sim \frac{9}{5}),  ~e=\frac{1}{5}\,(\sim \frac{6}{5}),~f= \frac{3}{5}, \label{eq:XABC}
}
where we used Eqs. (\ref{eq:Uones}) and (\ref{eq:YandX}) and used quantum numbers of  $ {\bf 35}=\Psi^{[ABC]} $. Because of the compact group nature, the naive U(1) charge calculation given in the bracket just by the tensor representation components is not exact. We use the $|X|\le 1$ for the fundamental representation Eq. (\ref{eq:YandX}). With two SU(5) indices, the $|X|$ charge are redundantly added, and we subtract $\pm 1$. With one more indices in addition to the two indices, again  we subtract $\pm 1$ once more. The rule to use in Eqs. (\ref{eq:XAB}) and (\ref{eq:XABC}) is to subtract $(N-1)$ from $X$ for $N$ SU(5) indices. Because $d=-\frac15$ and $e=\frac15$, one vectorlike pair of $\ten$ and $\tenb$ are removed at the GUT scale and we obtain two $\ten_{1/5}$'s from two {\bf 21} of Eq. (\ref{eq:XAB}) and one $\ten_{1/5}$ from  {\bf 35} of Eq. (\ref{eq:XABC}). In particular, note that $\tenb_{-1/5}$  of Eq. (\ref{eq:XABC}) contains $\overline{N}$ which can develop a VEV.  Thus, there result three SM families. Therefore, for the chiral representations we treat the anti-SU(5) representations as usual. For the BEH scalars, we need U(1) charges of the anti-SU(5) as $\five_{-2/5}$ which houses $H_d$ and $\fiveb_{+2/5}$ which houses $H_u$.

Now we can calculate the Yukawa coupling matrices for the quark sector. Here, we attempt to calculate $V_{\rm CKM}$, and comment on $U_{\rm PMNS}$ in the end. For charged leptons including $e^+,\mu^+,\tau^+$, which appear as SU(7) singlets, we must obtain all SU(7) singlet spectra. These singlets are not available at present. Thus, we try to calculate $V_{\rm CKM}$ and  $U_{\rm PMNS}$ without the knowledge on the singlets. The CKM matrix is obtained if we know the $Q_{\rm em}=\frac23$ and $\frac{-1}{3}$ quark mass matrices,
\dis{
&\overline{u}_LM^{(2/3)} u_R=\ten_{1/5}\,\fiveb_{-3/5}\, \langle \overline{\Phi}_{\rm BEH,2/5}\cdot(\cdots)\rangle
\\
&\overline{d}_L M^{(-1/3)}d_R=\ten_{1/5}\,\ten_{1/5}\, \langle  {\Phi}_{\rm BEH, -2/5}\cdot(\cdots)\rangle
}
where we used the anti-SU(5) notation. For the PMNS matrix, we do not need  information on the Yukawa couplings of the charged leptons. On the other hand, we need to know
\dis{
\ten_{1/5}\,\fiveb_{-3/5}\, \langle \overline{\Phi}_{\rm BEH,2/5}\cdot(\cdots)\rangle
}
for the Dirac mass of $N_i$ (the (45) element of $\ten$) and $\nu_j$ and $N_i -N_k $ masses.
The Dirac mass coupling is shown in Fig. (a) and the Majorana mass term of $N$ is shown in Fig. (b). The seesaw is a double seesaw as depicted in Fig 1\,(c), which is obtained from Fig. 1(a) and Fig. 1(b). But, we do not need the charged lepton mass matrices.

%%%%%%%%%%%%%%%%%%
\begin{figure}[!t]
\begin{center}
\includegraphics[width=0.65\linewidth]{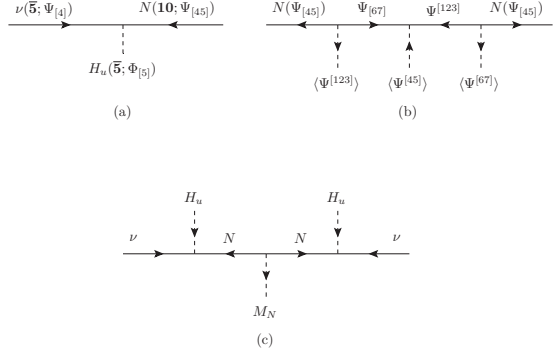}
\end{center}
\caption{The double seesaw with (a) The Dirac mass and (b) the Majorana mass of $N$, and (c) the seesaw mass of the SM neutrinos.} \label{fig:MajNuMass}
\end{figure}
%%%%%%%%%%%%%%%%%%%%%
 
As commented in Ref. \cite{KimUGUTF15}, the $b$-quark mass is expected to be much smaller than the $t$-quark mass, $ O( \langle T^{ \bf 21 }_{3,\rm BEH}\rangle \langle T^{ \bf 7 }_{3,\rm BEH} \rangle/ M_s \langle T^{ \bf 7 }_{6,\rm BEH}\rangle    )$, where $ \langle T^{ \bf 21 }_{3,\rm BEH}\rangle$ is the SU(5) splitting VEV $\langle\Phi^{[67]}\rangle$. Thus, we expect $m_b/m_t\sim\frac{\langle\Phi^{[67]}\rangle}{M_s\tan\beta}$. Even if $\tan\beta=O(1)$, we can fit $m_b/m_t$ to the observed value by appropriately tuning $\langle\Phi^{[67]}\rangle$. A similar suppression occurs for the second family members.

%%%%%%%%%%%%%%%%%%%%%%%%%%%%%%%%%%%%%
\subsection{A democratic submatrix of $M_{ \rm weak }$}\label{subsec:Submatrix}

  The multiplicity 2 of the fields from $T_3$ leads to a democratic form for the submatrix of the mass matrix. Thus, we consider
 \dis{
 \begin{pmatrix}
 \frac12& \frac12\\ \frac12&\frac12
 \end{pmatrix}  
 \to
 \begin{pmatrix}
 0& 0\\ 0&1
 \end{pmatrix}
}
 which can be diagonalized to give the eigenvalues $0$ and $1$.
The democratic form can be extended to have a permutation symmetric form $S_2$   
 which has only singlet representations. Introducing two small numbers $x$ and $y$ (for the two independent singlets) for breaking the $S_2$ symmetry, it can be diagonalized to
 \dis{
 M=\begin{pmatrix}
 \frac12 +\frac{y}{2},& \frac12+\frac{x}{2}\\[0.2em] \frac12+\frac{x}{2},&\frac12+\frac{y}{2}
 \end{pmatrix}  
 \to
 \begin{pmatrix}
  \frac{-x + y}{2},&0\\[0.2em] 0,&1+\frac{x+y}{2}
 \end{pmatrix}  \to
\begin{pmatrix}
 -\epsilon+\epsilon',& 0\\[0.2em] 0,&1+\epsilon
 \end{pmatrix},~{\rm with}~\epsilon=\frac{x+y}{2},~\epsilon'=y. \label{eq:diagDemo}
}
by
\dis{
U_{2\times 2}^\dagger MU_{2\times 2}, ~~{\rm with}~U_{2\times 2}=\begin{pmatrix}
\frac{1}{\sqrt2},&\frac{1}{\sqrt2}\\
-\frac{1}{\sqrt2},&\frac{1}{\sqrt2}\end{pmatrix}.\label{eq:twobytwoU}
}
A $3\times 3$ mass matrix is changed, using a $U_{3\times 3}$ matrix,  
\dis{
U_{3\times 3}^\dagger 
\begin{pmatrix} u_1,&u_2,&u_2\\[0.3em]
u_3^*,&\frac{1}{2}+\frac{y}{2},&\frac{1}{2}+\frac{x}{2}\\[0.3em]
u_3^*,&\frac{1}{2}+\frac{x}{2},&\frac{1}{2}+\frac{y}{2} 
\end{pmatrix}
U_{3\times 3}\to \begin{pmatrix} u_1,&0,&\epsilon_2\\[0.3em]
0,&-\frac{x}{2},&0\\[0.3em]
\epsilon_3^*,&0,&1+\frac{x}{2}
\end{pmatrix}
}
where $U_{3\times 3}$ contains the $U_{2\times 2}$ submatrix.
Here, $u$'s denote small parameters, breaking $S_2$ spontaneously  by the GUT scale VEVs of some SM singlet fields: $u=O(\langle \Phi\rangle/M_s)$. In view of the worry on the gravity spoil of discrete symmetries \cite{KraussWilczek,Ibanez92,Barr92,KimPRL13}, two singlet fields are better to be two components of a doublet representation $\Phi$ of a hypothetical gauge group SU(2) in the bottom-up scenario.\footnote{In the top-down scenario, there will be no gravity spoil problem, presumably satisfying the above condition automatically.} 
 The VEV $\langle \Phi\rangle$ breaks the $S_2$ symmetry spontaneously \cite{KimPRL13}. Then, the trace of $\Phi$ quantum number is zero. Thus, trace of Eq. (\ref{eq:diagDemo}) is 1, leading to $\epsilon'=0$. Thus, for the gravity-safe correction, which is our case arising from string compactification, let us diagonalize the democratic form to
 \dis{
\begin{pmatrix}
 -\frac{x}{2},& 0\\[0.2em] 0,&1+\frac{x}{2}
 \end{pmatrix}.
 }

%%%%%%%%%%%%%%%%%%%
Therefore, from the information on the origin of families in the untwisted  and twisted sectors ($U_1, T_3,T_5^+$) \cite{KimUGUTF15}, we can write the up- and down-type mass matrices as
   \dis{
\frac{M^{(u)}}{m_t}\approx  \begin{pmatrix}
 & |&\Psi_{[A]}(T_5^+)& \Psi_{[A]}(T_3) &\Psi_{[A] }(T_3)\\  
 -----&|&------&------&------\\
\Psi^{[ABC]} (U_1)&|&\epsilon_u&0 &\epsilon_2 \\[0.3em] 
 \Psi^{[AB]} (T_3)&|&0&x_c &0\\[0.3em]  
 \Psi^{[AB]} (T_3)&|&\epsilon_3^*&  0&    1
 \end{pmatrix} \label{eq:Mup}
 } 
  \dis{
\frac{M^{(d)}}{m_b}\approx  \begin{pmatrix}
 & |&\Psi^{[ABC]}(U_1)& \Psi^{[AB]}(T_3) &\Psi^{[AB] }(T_3)\\  
 -----&|&------&------&------\\
\Psi^{[ABC]} (U_1)&|&\epsilon_d &0  &\epsilon_1 \\[0.3em] 
 \Psi^{[AB]}(T_3) &|&0& x_s &0\\[0.3em]  
 \Psi^{[AB]}(T_3) &|&\epsilon_1 &0& 1 
 \end{pmatrix} \label{eq:Mdown}
 } 
 where the parameters in Eqs. (\ref{eq:Mup},\ref{eq:Mdown}) can be complex in general.  Note that $M^{(u)}$ is not a Hermitian matrix and $M^{(d)}$ is a symmetric matrix. In the bases where  Eqs. (\ref{eq:Mup}, \ref{eq:Mdown}) are written, we proceed to calculate the CKM and PMNS matrices. Parameter $\epsilon_{1}$ is given in the democratic form of the $2\times 2 $ matrix. But Eq. (\ref{eq:Mdown}) is written in the bases where the democratic form is broken. Thus, we expect two parameters $\epsilon_1(1\pm O(x_s))$. Since $x_s$ is small, we neglect this $S_2$ breaking correction. Similar comments apply to $\epsilon_2$ and $\epsilon_3$.
 
%%%%%%%%%%%%%%%%%
\subsection{The CKM matrix}

Since $M^{(d)}$ is symmetric, let us absorb two phases  $\epsilon_1$ and $\epsilon_d$ in $\Psi^{[AB]}(T_3)$ and $\Psi^{[ABC]}(T_3)$. So, the $d$-quark Yukawa couplings can be considered real. And we allow a real VEV for $H_d^0$. If it were complex, its phase can be absorbed to right-handed $d$ quarks. Then the real symmetric matrix  $M^{(d)}$ is diagonalized by an orthogonal matrix $O=O_L=O_R$,
\dis{
 M_{ \rm weak } ^{( d)}= O\begin{pmatrix}m_d&0&0\\ 0&m_s&0\\ 0&0& m_b\end{pmatrix}O^T 
}
where
\dis{
M_{ \rm weak } ^{( d)}= \begin{pmatrix}
\begin{array}{l}~~m_d   c_1^2 
 +m_s   c_2^2 s_1^2 \\
 ~~+m_b s_1^2s_2^2  
 \end{array},
 &\begin{array}{l}m_d c_1 c_3s_1\\
 - m_s c_2s_1[c_1c_2 c_3+ s_2s_3]\\ - m_b s_1s_2 [-c_2 s_3+c_1c_3 s_2]
 \end{array},
&\begin{array}{l} m_d c_1  s_1s_3\\
 - m_s  c_2 s_1 [c_1c_2s_3- s_2c_3]\\
 - m_b s_1s_2[c_2c_3+c_1s_2s_3]
\end{array}\\  \\
\begin{array}{l} m_d   c_1c_3 s_1\\
- m_s c_2s_1[c_1c_2c_3+ s_2s_3]\\
 - m_b s_1s_2[-c_2 s_3 +c_1s_2 c_3 ]   
\end{array},
&\begin{array}{l} m_d c_3^2s_1^2 \\
 + m_s [c_1c_2c_3+ s_2s_3]^2 \\
 + m_b  [-c_2s_3+c_1s_2c_3]^2
\end{array},
&\begin{array}{l} m_d c_3 s_1^2s_3\\
 + m_s [c_1c_2c_3+ s_2s_3 ]\\ ~~\cdot [c_1c_2s_3- s_2c_3 ]\\
  + m_b  [-c_2s_3+c_1s_2c_3]\\ ~~\cdot[c_2c_3+c_1s_2s_3 ]
\end{array}\\ \\
\begin{array}{l} m_d c_1s_1 s_3 \\
 - m_s  c_2s_1[c_1c_2s_3-s_2c_3] \\
 - m_b s_1s_2 [c_2 c_3 +c_1 s_2 s_3]   
\end{array},
&\begin{array}{l} m_d c_3s_1^2s_3 \\
 + m_s  [c_1c_2s_3-s_2c_3 ]\\ ~~\cdot[c_1c_2c_3+ s_2s_3 ]\\
  + m_b  [c_2c_3+c_1s_2s_3 ]\\ ~~\cdot [-c_2s_3+c_1s_2c_3 ]
\end{array} ,
&\begin{array}{l} m_d s_1^2s_3^2\\
 + m_s   [c_1c_2s_3- s_2c_3 ]^2\\
  + m_b   [c_2c_3+c_1s_2s_3 ]^2
\end{array}
\end{pmatrix}   \label{eq:MdWeak}
}
where $\theta_i$ represent the orthogonal matrix angles $\theta_{{\rm O},i}$.
Here, $O$ is taken as a real KS parametrization,
 \dis{
 V^{\rm KS}_{\rm real}&=\left(
\begin{array}{ccc}
c_{\rm O,1} & s_{\rm O,1}c_{\rm O,3} & s_{\rm O,1}s_{\rm O,3} \\
-c_{\rm O,2}s_{\rm O,1} &  s_{\rm O,2}s_{\rm O,3}+c_{\rm O,1}c_{\rm O,2}c_{\rm O,3} & - s_{\rm O,2}c_{\rm O,3}+c_{\rm O,1}c_{\rm O,2}s_{\rm O,3}  \\
- s_{\rm O,1}s_{\rm O,2} & -c_{\rm O,2}s_{\rm O,3}+c_{\rm O,1}s_{\rm O,2}c_{\rm O,3}  & c_{\rm O,2}c_{\rm O,3}+c_{\rm O,1}s_{\rm O,2}s_{\rm O,3} \\
\end{array}\right). \label{eq:KSCKM}
}
We consider $m_d=O(\lambda^4)\times m_b$. In Eq. (\ref{eq:MdWeak}), the (23) and (32) elements are vanishing up to O$(\lambda^9)$ for
\dis{
s_{\rm O,1}=0,~~t_{\rm O,2}= t_{\rm O,3}. \label{eq:Mddetermine}
}
where the angles are in the 1st quadrant.  Angles given in (\ref{eq:Mddetermine})
matches to Eq. (\ref{eq:Mdown}). Thus, there is one angle parameter in $V^{\rm KS}_{\rm real}$, which is taken as $\theta_{\rm O}=\theta_{\rm O,2}=\theta_{\rm O,2}$. So, the orthogonal matrix diagonalizing $M^{(d)}$ is
 \dis{
 V_L^{(d)}=V_R^{(d)} = \left(
\begin{array}{ccc}
1 & 0 & 0 \\
0 &  1 & 0  \\
0 &0 & 1 \\
\end{array}\right). \label{eq:KSCKM}
}
However, because of the $S_2$ breaking effect as commented above, $V_{L,R}^{(d)}$ contains small  parameters of $O(\epsilon_1 x_s)$. For simplicity, we neglect the  $O(\epsilon_1 x_s)$ correction. In the model of Ref. \cite{KimUGUTF15}, $\epsilon_1=O ( V_{\rm GUT}/M_s )$. This is because one may consider the following for $\epsilon_1$
\dis{
\frac{1}{M_s^2}\epsilon_{ABCDEFG}\Psi^{[ABC]}_{U_1}
\Psi^{[DE]}_{T_3}  \Phi^{F}_{T_3,\rm BEH} \langle \Phi^{G}_{T_3,\rm BEH} \rangle \langle \one_{T_{11},\rm BEH} \rangle,\nonumber
}
and $m_b=O ( V_{\rm GUT}/M_s )$.  Thus, $\epsilon_1 x_s$ is estimated to be O($\lambda^{4}$).
Then,  the determination of the CKM matrix depends  approximately on the diagonalization of  $M^{(u)}$.
 
%%%%%%%%%%%%%%%%%
\subsection{The CKM and PMNS matrices from anti-SU(7) UGUTF}\label{subsec:CKMfromUGUTF}
  
Now the CKM matrix is determined from the diagonalization of $M^{(u)}$ by bi-unitary matrices: $V_L^{(u)}$ and $V_R^{(u)}$ with $V_L^{(u)}\ne V_R^{(u)}$,
\dis{
V_{\rm CKM} =V_L^{(u)}  O_L^{(d)\,T}\simeq V_L^{(u)}
  }  
which does not depend on $V_R^{(u)}$.
The matrix elements and $\Qem=\frac23$
   quark masses have the following relations
   \dis{
&{u}^{({\rm mass}\,i)}_L= V^{ia}_L u^a_L, ~~{u}^{({\rm mass}\,i)}_R= V^{ia}_R u^a_R, \\
&\bar{u}^b_R M^{ba}_{{\rm weak},u} u^a_L= \bar{u}^{({\rm mass}\,j)}_R(V_R)^{jb} M^{ba}_{{\rm weak},u}  (V^\dagger_L )^{ai}u^{({\rm mass}\,i)}_L.
}
Thus, the mass matrix elements in the weak basis are
\dis{
M^{ba}_{{\rm weak},u} =(V^\dagger_R)^{bj}  M ^{ji}_{{\rm diag},u} (V_L)^{ia} =  m_u (V^\dagger_R)^{b1} (V_L)^{1a} 
 + m_c (V^\dagger_R)^{b2}(V_L)^{2a} + m_t (V^\dagger_R)^{b3} (V_L)^{3a},
}
or
\dis{
\begin{pmatrix}
\begin{array}{l}~~m_u c_1' c_1\\
 +m_c  c_2's_1'c_2s_1 \\
 +m_t s_1's_2'  s_1s_2 e^{i\delq-i\delq'}
 \end{array},
 &\begin{array}{l}m_u c_1' s_1c_3\\
 - m_c c_2's_1'[c_1c_2c_3+ s_2s_3e^{-i\delq}]\\ - m_ts_1's_2'e^{-i\delq'}\\ ~~\cdot [-c_2s_3+c_1s_2c_3e^{i\delq}]
 \end{array},
&\begin{array}{l} m_u c_1' s_1s_3\\
 - m_c  c_2's_1'[c_1c_2s_3- s_2c_3e^{-i\delq}]\\
  - m_t s_1's_2'e^{-i\delq'}\\ ~~\cdot[c_2c_3+c_1s_2s_3e^{i\delq}]
\end{array}\\  \\
\begin{array}{l} m_u c_3's_1' c_1\\
- m_c [c_1'c_2'c_3'+ s_2's_3'e^{i\delq'}]c_2s_1\\
 - m_t [-c_2's_3'+c_1's_2'c_3'e^{- i\delq'}]\\ ~~\cdot s_1s_2e^{i\delq}
\end{array},
&\begin{array}{l} m_u c_3's_1' s_1c_3\\
 + m_c [c_1'c_2'c_3'+ s_2's_3'e^{i\delq'}]\\ ~~\cdot[c_1c_2c_3+ s_2s_3e^{-i\delq}]\\
 + m_t  [-c_2's_3'+c_1's_2'c_3'e^{- i\delq'}]\\ 
~~ \cdot[-c_2s_3+c_1s_2c_3e^{i\delq}]
\end{array},
&\begin{array}{l} m_u c_3's_1' s_1s_3\\
 + m_c [c_1'c_2'c_3'+ s_2's_3'e^{i\delq'}]\\ ~~\cdot [c_1c_2s_3- s_2c_3e^{-i\delq}]\\
  + m_t  [-c_2's_3'+c_1's_2'c_3'e^{- i\delq'}]\\ ~~\cdot[c_2c_3+c_1s_2s_3e^{i\delq}]
\end{array}\\ \\
\begin{array}{l} m_us_1's_3'c_1\\
 - m_c  [c_1'c_2's_3'-s_2'c_3'e^{ i\delq}]c_2s_1 \\
 - m_t  [c_2'c_3'+c_1's_2's_3'e^{-i\delq'}]\\ ~~\cdot s_1s_2e^{i\delq}
\end{array},
&\begin{array}{l} m_u s_1's_3' s_1c_3\\
 + m_c  [c_1'c_2's_3'-s_2'c_3'e^{ i\delq'}]\\ ~~\cdot[c_1c_2c_3+ s_2s_3e^{-i\delq}]\\
  + m_t  [c_2'c_3'+c_1's_2's_3'e^{-i\delq'}]\\ ~~\cdot [-c_2s_3+c_1s_2c_3e^{i\delq}]
\end{array} ,
&\begin{array}{l} m_u s_1's_3' s_1s_3\\
 + m_c  [c_1'c_2's_3'-s_2'c_3'e^{ i\delq}]\\ ~~\cdot[c_1c_2s_3- s_2c_3e^{-i\delq}]\\
  + m_t [c_2'c_3'+c_1's_2's_3'e^{-i\delq'}]\\ ~~\cdot [c_2c_3+c_1s_2s_3e^{i\delq}]
\end{array}
\end{pmatrix} \label{eq:UmassinW}  
}   
where angles in $V_L^{(u)}$ are $\theta_i,\delta$ and angles in $V_R^{(u)}$ are $\theta_i',\delta'$. 

Comparing Eqs. (\ref{eq:Mup}) and (\ref{eq:UmassinW}), we have 9 constraints.
The order of magnitudes of the elements are such that the determinant of mass matrix is O($\lambda^8m_t^3$) with $m_c=O(\lambda^2)m_t$ and $m_u=O(\lambda^6)m_t$. Thus, the product of (11), (23), and (32) elements is O($\lambda^8m_t^3$), and the product of (11), (22), and (33) elements is also O($\lambda^8m_t^3$). So, let the (22) element is O($\lambda m_t $) or  O($\lambda^2 m_t $).  
For  $M_{(11)}^{(u)}$=O($\lambda^5 m_t $),  we require  $M_{(23)}^{(u)}$= O($\lambda^{3/2} m_t $) and
$M_{(32)}^{(u)}$=O($\lambda^{3/2} m_t $).
For $M_{(11)}^{(u)}$=O($\lambda^4 m_t $),
we require   $M_{(23)}^{(u)}$=O($\lambda^2 m_t $) and$M_{(32)}^{(u)}$=O($ \lambda^2 m_t $). So, whether the (22) element is O($\lambda m_t $) or  O($\lambda^2 m_t $), we consider$M_{(23)}^{(u)}$=O($\lambda^2 m_t $) and $M_{(32)}^{(u)}$=O($ \lambda^2 m_t $). Because the (33) element is O(1), we require both (12) and (21) elements to be O($\lambda^4$). By the same argument, we require both (13)  and (31) elements to be O($\lambda^3$).  Thus, we require 
\begin{eqnarray}   
&(11)&\lesssim O(\lambda^4)m_t\label{eq:uu}\\
&(12)&\lesssim O(\lambda^4)m_t  \\
&(13)& =O(\lambda^3)m_t,  \label{eq:ut}\\
&(21)&\lesssim O(\lambda^4)m_t  \\
&(31)& =O(\lambda^3)m_t,  \label{eq:cu}\\
&(22)&\lesssim O(\lambda^2~{\rm or}~\lambda)m_t  \label{eq:cc} \\
&(23)&=O(\lambda^2)m_t   \label{eq:ct} \\
&(32)&=O(\lambda^2)m_t \label{eq:tc}\\
&(33)&=O(1)m_t\label{eq:tt}
\end{eqnarray}  
 where we used $m_t=173.21 ,m_c=1.275$, and $\lambda=\sin\theta_1\cos\theta_3=0.2253$. The determinant can be  $m_um_cm_t$ with (31), (22), and (13) elements for the orders given above. So, we take (11), (12), and (21) elements with inequality signs.
  
  Before presenting a numerical study, let us check that solutions suggested in Eqs. (\ref{eq:uu}--\ref{eq:tt}) are possible.
  From the (23) element, we restrict $s_2'$ and $s_3'$ at order $\lambda^2$,
   \dis{
     (23):&   m_u c_3's_1' s_1s_3 
 + m_c [c_1'c_2'c_3'+ s_2's_3'e^{i\delq'}]   \cdot [c_1c_2s_3- s_2c_3e^{-i\delq}]\\
  &+ m_t  [-c_2's_3'+c_1's_2'c_3'e^{- i\delq'}]   \cdot[c_2c_3+c_1s_2s_3e^{i\delq}]\simeq 0\\
  &\to s_2'=O(\lambda^2),s_3'=O(\lambda^2).
  \label{eq:s2prime}
  }
Then, we satisfy (32) and (22) elements,
  \dis{ 
(32):&  m_u s_1's_3' s_1c_3 
 + m_c  [c_1'c_2's_3'-s_2'c_3'e^{ i\delq'}]   \cdot[c_1c_2c_3+ s_2s_3e^{-i\delq}]\\
  &+ m_t  [c_2'c_3'+c_1's_2's_3'e^{-i\delq'}]   \cdot [-c_2s_3+c_1s_2c_3e^{i\delq}]=O(\lambda^2) ,    }
\dis{
(22): &m_u c_3's_1' s_1c_3 
 + m_c [c_1'c_2'c_3'+ s_2's_3'e^{i\delq'}]   \cdot[c_1c_2c_3+ s_2s_3e^{-i\delq}]\\
 &+ m_t  [-c_2's_3'+c_1's_2'c_3'e^{- i\delq'}] 
 \cdot[-c_2s_3+c_1s_2c_3e^{i\delq}]=O(\lambda^2).
}
Now,  the (12) element restricts $s_1'$ at order $\lambda^2$,
\dis{
(12):& m_u c_1' s_1c_3 
 - m_c c_2's_1'[c_1c_2c_3+ s_2s_3e^{-i\delq}]  - m_ts_1's_2'e^{-i\delq'} \cdot [-c_2s_3+c_1s_2c_3e^{i\delq}]=O(\lambda^4)\\
 &\to s_1'=O(\lambda^2) \label{eq:s1prime}
  }  
Then, the (11) element is very small,  O($\lambda^5$).
The remaining (21), (13), and (31)   elements are
\begin{eqnarray}
(21):&m_u c_3's_1' c_1 
- m_c [c_1'c_2'c_3'+ s_2's_3'e^{i\delq'}]c_2s_1 
 - m_t [-c_2's_3'+c_1's_2'c_3'e^{- i\delq'}]   \cdot s_1s_2e^{i\delq}=O(\lambda^3),\label{eq:element21}
\\
(13):&m_u c_1' s_1s_3
 - m_c  c_2's_1'[c_1c_2s_3- s_2c_3e^{-i\delq}]   - m_t s_1's_2'e^{-i\delq'} \cdot[c_2c_3+c_1s_2s_3e^{i\delq}]
 =O(\lambda^4) ,\\
(31):& m_us_1's_3'c_1 
 - m_c  [c_1'c_2's_3'-s_2'c_3'e^{ i\delq}]c_2s_1  - m_t  [c_2'c_3'+c_1's_2's_3'e^{-i\delq'}] \cdot s_1s_2e^{i\delq}=O(\lambda^3) \label{eq:element31}
\end{eqnarray}
where we considered $m_c=O( \lambda^2)m_t$. Here the rough bound of Eqs. (\ref{eq:uu}--\ref{eq:tt}) are satisfied except in Eq.  (\ref{eq:element21}). But,  $m_c$ is between $O( \lambda^2)m_t$ and $O( \lambda^3)m_t$ and Eqs.  (\ref{eq:element21}) is acceptable in our rough estimation. In our order of estimation, $\delq'$ is not  restricted.\footnote{In the numerical study below, $\theta_3'$ is not bounded also.}
    
Therefore, the mass matrices Eqs. (\ref{eq:Mup}) and (\ref{eq:Mdown}) obtained from anti-SU(7) UGUTF leads to a reasonable CKM matrix. Similarly, one can consider the lepton mixing angles which however need singlet contributions. Since there will appear additional parameters for the unkown heavy neutral lepton masses, there will be more freedom fitting for a reasonable PMNS matrix \cite{PMNS}. 

%%%%%%%%%%%%%%%%
\section{Bounds on the parameters of right-handed unitary matrix $V_R^{(u)}$}\label{sec:NumBounds}

%%%%%%%%%%%%%%%%%%
\begin{figure}[!t]
\begin{center}
\includegraphics[width=0.7\linewidth]{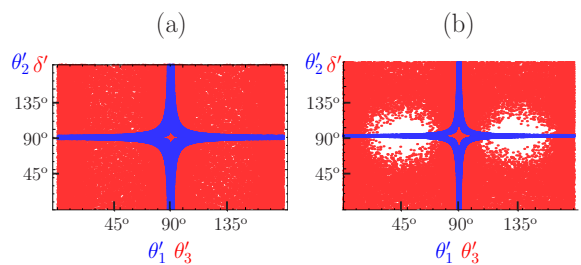} 
\end{center}
\caption{The bounds on the angles of the right-handed unitary matrix $V_R^{(u)}$ diagonalizing $M^{(u)}$, (a) $a=\frac{1}{1.5}\sin\theta_c,~b=1.5\sin\theta_c$, and (b) $a=\frac{1}{1.2}\sin\theta_C,~b=1.2\sin\theta_c$.   The white regions are not allowed.} \label{fig:Rangles}
\end{figure}
%%%%%%%%%%%%%%%%%%%%%
 
In  Fig. \ref{fig:Rangles}, we present the allowed angles of  $V_R^{(u)}$. The color code is: the projection on $\theta_2'$ versus  $\theta_1'$ for all allowed $\theta_3'$ and $\delq'$ as blue, and  $\delq'$  versus  $\theta_3'$ for all allwed $\theta_1'$ and $\theta_2'$ as red. We allowed the 1\,$\sigma$ for $\theta_1,\theta_2,\theta_3$ and $\delq$ in $V_L^{(u)}$. We choose the $V_L$ angles as  $\theta_1= 13.025^{{\rm o\,}{+0.039}^{\,\rm o}}_{~\,-0.038^{\,\rm o}},  
\theta_2=  2.292^{{\rm o\,}{+2.625}^{\,\rm o}}_{~\,-2.217^{\,\rm o}},$\footnote{The lower limit is given from the measured value of $J\simeq 10^{-5}$.}
$
\theta_3=8.8923^{{\rm o\,}{+0.0382}^{\,\rm o}}_{~\,-0.0357^{\,\rm o}} $, and $\delq=85.4^{{\rm o\,}{+3.9}^{\,\rm o}}_{~\,-3.8^{\,\rm o}}$ \cite{CKMPDG15}. For the right-hand sides (of equality or inequalities) in Eqs. (\ref{eq:uu}--\ref{eq:uu}), the expansion parameter $\lambda^n$ is varied in the region $a^n\le \lambda^n\le b^n$.  In   Fig. \ref{fig:Rangles} (a), we choose $a=\frac23\sin\theta_C$ and $b= \frac32\sin\theta_C$.  In   Fig. \ref{fig:Rangles} (b), we choose $a=\frac1{1.2}\sin\theta_C$ and $b= 1.2\sin\theta_C$.   From Fig.  \ref{fig:Rangles}, we conclude that the mass matrices  Eqs. (\ref{eq:Mup}) and (\ref{eq:Mdown}), suggested from the UGUTF anti-SU(7), are phenomenologically allowed.

%%%%%%%%%%%%%%%%
\section{Conclusion}\label{sec:Conclusion}
 We presented bounds on the mixing angles of the right-handed currents, diagonalizing the quark mass matrices, suggested from a recently proposed families unified GUT model based on anti-SU(7) \cite{KimUGUTF15}. The investigation suggests that quark mass matrices suggested in \cite{KimUGUTF15} are phenomenologically allowable, and a numerical search is presented in figures on four mixing angles of $V_R^{(u)}$ within the $1\sigma$ bounds of the CKM parameters, $\theta_1,\theta_2,\theta_3,$ and $\delq$. Along the way, we also commented on some aspects of the Jarlskog determinant. In particular, the currently allowed CKM parametrization falls into three classes for choosing $\delq=\alpha,\beta,$ or $\gamma$ of the PDG book. The Kobayashi-Maskawa and Kim-Seo parametrization choose $\delq=\alpha$ and Chau-Keung-Maiani parametrization chooses $\delq=\gamma$. It suggests that with three real CKM angles fixed, the Jarlskog determinant is maximum with $\alpha=\frac{\pi}{2}$.

 %%%%%%%%%%%%%%%%%%%%%%%%%%%%%%%%%%%%%%%%%%%%%%%%%%%%%%%%%%%%%%%%%%%%
\acknowledgments{We would like to thank Hyung Do Kim for useful communications. This work is supported in part by the National Research Foundation (NRF) grant funded by the Korean Government (MEST) (No. 2005-0093841) and by the IBS (IBS-R017-D1-2014-a00). }

%%%%%%%%%%%%%%%%%%
%%%%%%%%%%%%%%

  %%%%%%%%%%%%%%%%%%  

\end{document}